\documentclass[%
prb,
superscriptaddress,
nofootinbib,
 amsmath,amssymb,
 aps,
 twocolumn,
]{revtex4-1}
\usepackage{soul}
\usepackage[unicode=true,pdfusetitle,hypertexnames=false,
 bookmarks=true,bookmarksnumbered=false,bookmarksopen=false,
 breaklinks=true,pdfborder={0 0 0},pdfborderstyle={},backref=false,colorlinks=true]
 {hyperref}
\usepackage[T1]{fontenc}
\usepackage[justification=justified, format=plain]{caption}
\usepackage{ulem}
\usepackage{braket}
\usepackage{subfig}
\usepackage{graphicx}
\usepackage{dcolumn}
\usepackage{bm}

\begin{document}

\title{Unraveling correlated material properties with noisy quantum computers: Natural orbitalized variational quantum eigensolving \\ of extended impurity models within a slave-boson approach}

\author{P. Besserve}
\affiliation{Atos Quantum Laboratory, Les Clayes-sous-Bois, France
}
 \affiliation{Centre de Physique Th\'eorique, CNRS, Ecole Polytechnique, Institut Polytechnique de Paris, Palaiseau, France}
\author{T. Ayral}
\affiliation{Atos Quantum Laboratory, Les Clayes-sous-Bois, France
}

\date{\today}

\begin{abstract}
We propose a method for computing space-resolved correlation properties of the two-dimensional Hubbard model within a quantum-classical embedding strategy that uses a Noisy, Intermediate Scale Quantum (NISQ) computer to solve the embedded model.
While previous approaches were limited to purely local, one-impurity embedded models, requiring at most four qubits and relatively shallow circuits, we solve a two-impurity model requiring eight qubits with an advanced hybrid scheme on top of the Variational Quantum Eigensolver algorithm.
This iterative scheme, dubbed Natural Orbitalization (NOization), gradually transforms the single-particle basis to the approximate Natural-Orbital basis, in which the ground state can be minimally expressed, at the cost of measuring the one-particle reduced density-matrix of the embedded problem.
We show that this transformation tends to make the variational optimization of  existing (but too deep) ansatz circuits faster and more accurate, and we propose an ansatz, the Multireference Excitation Preserving (MREP) ansatz, that achieves great expressivity without requiring  a prohibitive gate count, thus bridging the gap between hardware-efficient and physically-motivated strategies in variational ansatz design.
The one-impurity version of the ansatz has only one parameter, making the ground state preparation a trivial step, which supports the optimal character of our approach.
Within a Rotationally Invariant Slave Boson embedding scheme that requires a minimal number of bath sites and does not require computing the full Green's function, the NOization combined with the MREP ansatz allow us to compute accurate, space-resolved quasiparticle weights and static self-energies for the Hubbard model even in the presence of noise levels representative of current NISQ processors.
This paves the way to a controlled solution of the Hubbard model with larger and larger embedded problems solved by quantum computers.

\end{abstract}

\maketitle

Strongly correlated materials and their complex phase diagrams---with the corresponding technological promises---still pose a great theoretical challenge: the many-body, or "ultra-quantum", nature of the underlying physics makes for an exponential difficulty in solving even one of the most simple models for describing such phenomena, the Hubbard model
\begin{equation}
    H_{\mathrm{Hub}} = t \sum \limits_{\langle i, j \rangle, \sigma}  c^{\dagger}_{i\sigma}c_{j\sigma} + U\sum \limits_i n_{i\uparrow} n_{i\downarrow},\label{eq:Hubbard}
\end{equation}
where $t$ and $U$ denote respectively the nearest-neighbor ($\langle i,j \rangle$) tunneling amplitude and on-site interaction of fermions of spin $\sigma=\uparrow,\downarrow$ on a lattice with sites $i=1\dots N$ (with $N \rightarrow \infty$).
The exponential scaling $4^N$ of the Hamiltonian size, or the exponentially small Monte-Carlo sign, have so far stymied attempts at finding a complete solution of the model for regimes of physical interest using classical computers.

\begin{figure}
  \centering
  \includegraphics[width=1.0\columnwidth]{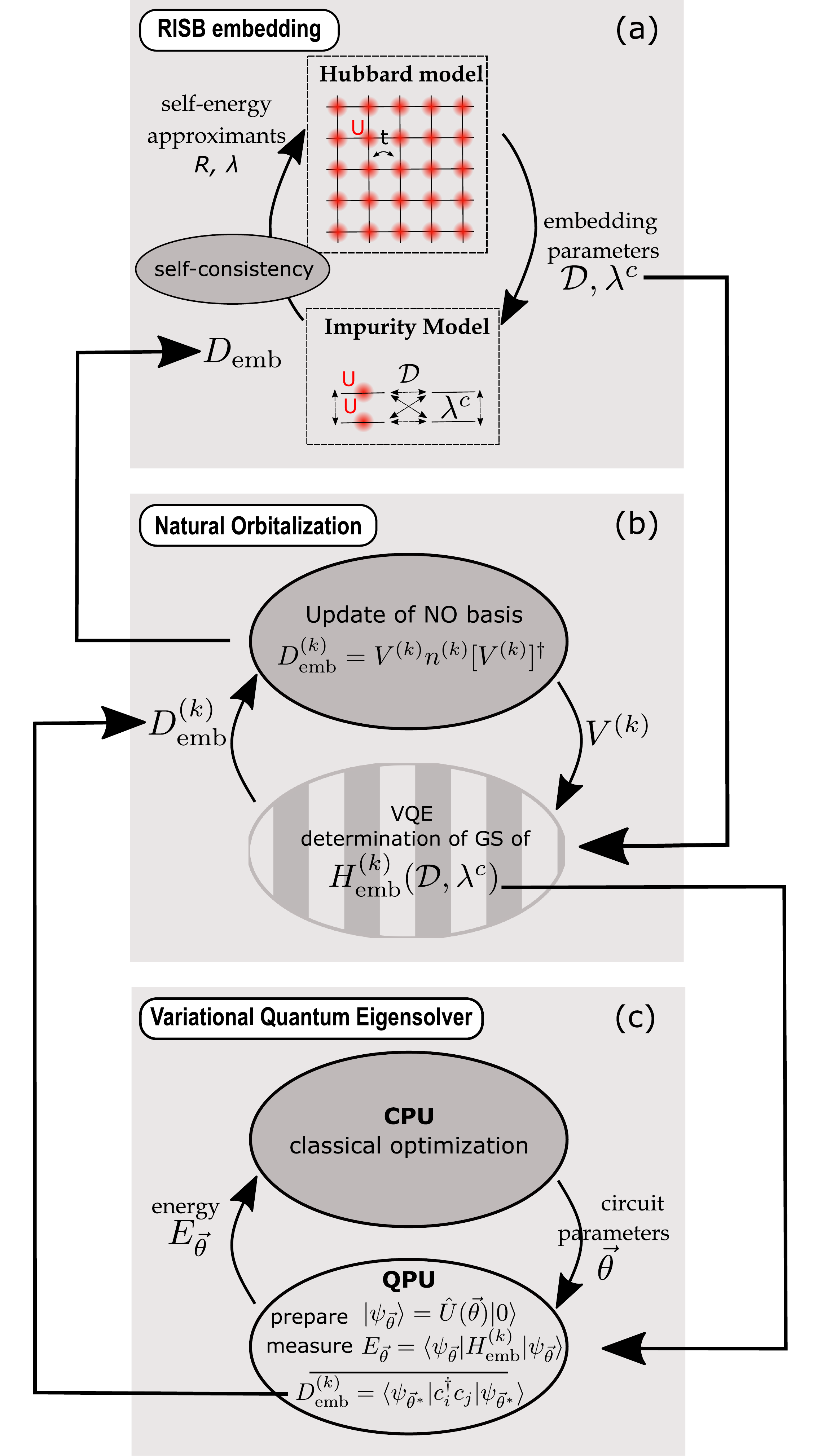}
  \caption{\textbf{Quantum-classical embedding scheme.}
  {\it (a)}: Mapping from the Hubbard model to an impurity model $H_\mathrm{emb}$.
  {\it (b)}: NOization: Iterative rotation to approximate NO basis via 1-RDM diagonalization.
  {\it (c)}: VQE solution of $H_\mathrm{emb}$ in current approximate NO basis.}\label{fig:method}
\end{figure}

Yet, recent advances in quantum computing technologies have raised expectations that quantum processors could be used to help remove, or at least lower, this exponential hurdle.
Early studies \cite{ Bauer2016,Kreula2016,Kreula2016a,Rubin2016} acknowledged the necessity of not tackling directly the full lattice problem with a quantum processing unit (QPU), but instead of resorting to hybrid quantum-classical approaches to boil the Hubbard model down to its quantum quintessence. 
In practice, the proposed methodology used well-known embedding strategies, like Dynamical Mean Field Theory (DMFT \cite{georges_dynamical_1996}) or Density Matrix Embedding Theory (DMET \cite{knizia_density_2012}), to map the Hubbard model to a so-called impurity or embedded model only comprising a few ($N_c$) correlated sites hybridizing with a fermionic bath, thereby reducing the dimension of the problem to be solved, while keeping its most relevant many-body features (Fig.~\ref{fig:method}(a), upper panel).

These early proposals, most of which assumed perfect QPUs, were followed by studies investigating the behavior of the method when using noisy QPUs, whether classically simulated\cite{jaderberg_minimum_2020} or actual\cite{keen_quantum-classical_2020,rungger_dynamical_2020,yao_gutzwiller_2021,tilly_reduced_2021} processors. 
In these studies, the noisy character of current devices, dubbed "Noisy, Intermediate Scale Quantum" (NISQ) processors \cite{Preskill2018}, led to another kind of quantum-classical hybridization: instead of solving the impurity model exclusively with one coherent quantum evolution (out of reach of NISQ QPUs), the workload was divided between the classical processor (CPU) and the QPU by making use of a now widespread variational method called Variational Quantum Eigensolving (VQE\cite{peruzzo_variational_2014}).
The freedom in choosing the variational state, or ansatz, to be prepared on the QPU, allowed to pick quantum circuits short enough that they could produce meaningful results before too many errors occurred.
Yet, these early studies were limited to (i) small impurity models (in terms of the number $N_c$ of correlated sites) or equivalently a small number of qubits (four at most) and (ii), for most \cite{jaderberg_minimum_2020,keen_quantum-classical_2020,rungger_dynamical_2020}, simplistic DMFT schemes ("two-site DMFT" \cite{Potthoff2001}); (iii) lastly, no systematic recipe was proposed to design ansatz circuits tailored to such impurity models.

In this work, we propose, implement and test a NISQ-compatible strategy towards increasing the size $N_c$ of the impurity model that can be tackled by a QPU, a crucial step to achieve a {\it controlled} solution of the Hubbard model using embedding methods\cite{lee_rotationally_2019}.
We introduce a novel quantum-classical hybrid strategy on top of the VQE, dubbed NOization, that iteratively looks for the orbital (or qubit) basis that systematically maximizes the representativity of a given ansatz.
This allows us to solve impurity models of size $N_c=2$ in the presence of realistic noise levels, paving the way for a spatial resolution of correlation effects, going beyond prior works that were limited to $N_c=1$.

\begin{figure}
    \centering
    \includegraphics[width=1.0\columnwidth]{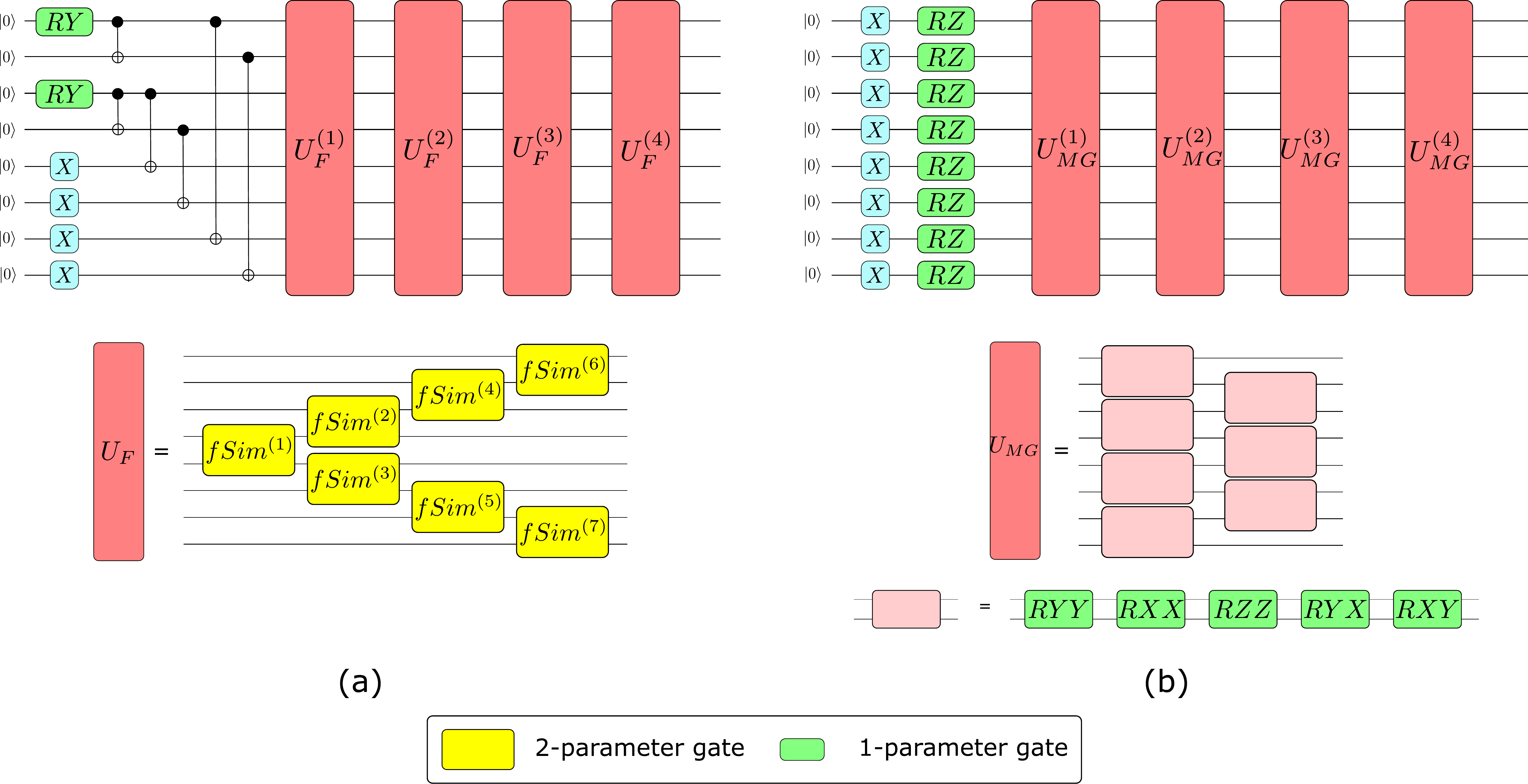}
    \caption{\textbf{Ansatz circuits.} (a) Top: Multireference Excitation Preserving (MREP) ansatz (4 layers, 58 angles). Bottom: breakdown of $U_F$ block. \textit{fSim} refers to Google's excitation-preserving fermionic simulation gate.
    (b) Top: Low-Depth Circuit Ansatz (LDCA) circuit (4 layers [=1 cycle], 148 parameters). Bottom: breakdown of $U_{MG}$ matchgates block.}
    \label{fig:circuits}
\end{figure}

{\it Method.} We solve the Hubbard model (Eq.~\eqref{eq:Hubbard}) using an embedding method called the Rotationally-Invariant Slave Boson (RISB \cite{lechermann_rotationally-invariant_2007}) method.
It self-consistently maps the Hubbard model onto a quantum impurity model
\begin{align}
  &\hat{H}_{\mathrm{emb}}[\mathcal{D}, \lambda^c] = 
  U \sum_{i} n_{i\uparrow} n_{i\downarrow} + \sum_{ij,\sigma} t_{ij} c^\dagger_{i\sigma} c_{j\sigma}\nonumber\\
  &+\sum \limits_{ab, \sigma} \left[\lambda^c\right]_{ab} f_{b\sigma} f^{\dagger}_{a\sigma} + \sum \limits_{ia, \sigma}(\left[\mathcal{D}\right]_{i a}c^{\dagger}_{i\sigma}f_{a\sigma} + \mathrm{h.c.})
  \label{eq:Hemb}
\end{align}
describing $2N_c$ (spinful) correlated orbitals (denoted by creation operators $c_{i\sigma}^\dagger$) hybridized with $2N_c$ uncorrelated "bath" orbitals (denoted by creation operators $f_{a\sigma}^\dagger$). It must be solved for its one-particle reduced density matrix (1-RDM) elements
$N_{ab} = \langle \Phi | f_{b\sigma} f_{a \sigma}^\dagger | \Phi \rangle$ and $M_{ia} = \langle \Phi | c_{i\sigma}^\dagger f_{a \sigma} | \Phi \rangle$, with $\Phi$ the ground-state wavefunction of $H_\mathrm{emb}$. 

The computational bottleneck of RISB is the computation of the ground-state wavefunction $|\Phi\rangle$ of the impurity model $H_\mathrm{emb}$. Despite its reduced dimension compared to $H_\mathrm{Hub}$, it still is a correlated many-body problem whose solution becomes exponentially difficult to compute with a classical computer as $N_c$ increases.
We are thus going to execute quantum circuits on quantum processors to solve this impurity model.
Before this, let us note that RISB has two complementary advantages in terms of the widths and depths of the requisite quantum circuits compared to the more accurate DMFT:
(i) RISB's impurity model has a finite number of bath sites that equals the number of correlated orbitals \cite{lanata_phase_2015}. This construction obviates the need for the arbitrary truncations that are required when projecting DMFT's hybridization function onto a finite number of bath sites, and, more importantly, caps the total number of required qubits (width) to $4 N_c$;
(ii) RISB does not require the computation of the full Green's function, avoiding deep Trotterization circuits \cite{Bauer2016}.
The price to pay for these lowered requirements is that RISB does not give access to the full frequency-dependent self-energy but to a low-energy expansion parameterized as:
\begin{equation}
    \Sigma(\omega) = \omega \left(1-(R^{\dagger}R)^{-1}\right) + R^{-1} \lambda (R^{\dagger})^{-1}.\label{eq:selfenergy}
\end{equation}
The two (matrix-valued) coefficients $R$ and $\lambda$---which give access to the quasiparticle renormalization $Z=R^\dagger R$ and static self-energy shift $R^{-1} \lambda (R^{\dagger})^{-1}$---as well as the bath energies $\lambda^c$ and hybridizations $\mathcal{D}$, are determined self-consistently (see Appendix~\ref{sec:RISB}).

To find an approximate ground state using NISQ devices, the usual method is VQE (Fig.~\ref{fig:method}(c)).
It consists in finding the parameters $\vec{\theta}^*$ of a family of ansatz states $\ket{\psi(\Vec{\theta})}$ that minimize the average energy $\langle \psi(\Vec{\theta}) | \hat{H}_{\mathrm{emb}} |\psi(\Vec{\theta})\rangle$. The parametric states are constructed via a quantum circuit $\ket{\psi(\Vec{\theta})} = \hat{U}(\theta_p)\cdots\hat{U}(\theta_2) \hat{U}(\theta_1) \ket{\emptyset}$, and their energy is measured by decomposing $H_\mathrm{emb}$ as a weighted sum of Pauli operators (that are measured 
on the QPU).
The minimization procedure is carried out by a classical processor using standard optimization algorithms (see Appendix~ \ref{subsec:classical_optim} for details).

The first challenge one must overcome to successfully implement a VQE procedure on NISQ hardware lies in the design of the ansatz circuit.
It faces two contradictory requirements: while it must be deep enough to capture a sufficiently representative portion of the Hilbert space, it must remain shallow enough that it does not suffer substantially from quantum errors.
These requirements are reflected in the two main approaches that have been taken to design VQE ansätze: the Hardware-Efficient Ansatz (HEA\cite{kandala_hardware-efficient_2017}, used, e.g., in Ref.~\onlinecite{keen_quantum-classical_2020}) route strives to fulfill the second requirement.
The first requirement is usually fulfilled by approaches inspired by prior knowledge on the problem at hand, like the Unitary Coupled-Cluster (UCC) ansatz, known to be an expressive (albeit deep) ansatz for quantum chemistry problems (used, e.g., in Ref.~\onlinecite{yao_gutzwiller_2021}), or the more recent Low-Depth Circuit Ansatz (LDCA \cite{dallaire-demers_low-depth_2019}), which essentially generates non-gaussian (i.e. correlated) wavefunctions starting from a circuit meant to generate gaussian (mean-field) states. 

While HEA approaches have been shown to work reasonably well for the smallest impurity models ($N_c=1$), they are, by essence, not well suited to a systematic extension to larger impurity problems.
Conversely, physically-motivated ansätze like LDCA can be systematically grown to tackle larger dimensional problems, but this comes with the major drawback of depths that quickly exceed the typical coherence times of NISQ QPUs. 
We have confirmed these intuitions through numerical simulations of noisy QPUs with NISQ noise levels: 
even the simplest instance of the LDCA circuit (comprising only one cycle) is ruled out by noise, whereas straightforwardly stacking HEA quantum routines such as 'thinly-dressed CNOT' gates \cite{jaderberg_minimum_2020} to get an ansatz able to prepare $N_c=2$ embedded models ground states requires far too many layers.

To remedy the above shortcomings, we use a hitherto oft-neglected degree of freedom, namely the single-particle orbital basis in which the problem is described.
While the Hubbard model and the impurity model have been expressed in the site-spin basis $(i,\sigma)$ (and $(a, \sigma)$), it is not the only choice.  
The orbital basis can be modified to optimize a given criterion.
For instance, the Orbital-Optimized VQE (OO-VQE\cite{mizukami_orbital_2020, sokolov_quantum_2020}) algorithm dresses the ubiquitous Unitary Coupled Cluster with single and double excitations (UCCSD) ansatz by an orbital basis rotation, whose parameters are optimized alongside the ones of the circuit to minimize the variational energy. Through this rotation, the UCC ansatz may contain excitations from other reference states than the Hartree-Fock state. It was also proposed \cite{koridon_orbital_2021} that orbital optimization be carried out as a classical preprocessing step in which the 1-norm of the dressed Hamiltonian, related to the magnitude of the coefficients of its different terms, is reduced. Such methods typically incur a large number of additional variational parameters to be tuned.

On the other hand, a well-known basis in quantum chemistry is the natural-orbital (NO) basis \cite{lowdin_quantum_1955}, which is defined as the single-particle basis that diagonalizes the 1-RDM associated with the exact ground state $|\Phi \rangle$.
It is the basis where $|\Phi \rangle$ can be written with the lowest number of Slater determinants. Intuitively, this means that this is the basis where the circuit to prepare $|\Phi \rangle$ will be the simplest.
Conversely, one can expect that for a given (fixed) ansatz circuit, putting the qubits in the NO basis will lead to the most representative state.

We therefore propose a method, dubbed NOization (shorthand for "Natural-Orbitalization"), that leverages the NO basis for maximizing the expressivity of a given ansatz, thereby enhancing the expressive power of even shallow (and therefore noise-robust) ansätze.
In order to solve the issue that the exact ground state $|\Phi \rangle $ needed to compute the NO basis is unknown, we introduce an iterative algorithm similar in spirit to the purely classical methods proposed in Refs~\onlinecite{lu_efficient_2014,lu_natural-orbital_2019}, which gradually transforms the single-particle basis to an approximate NO basis as follows (Fig.~\ref{fig:method}(b)): starting from a given guess $|\psi(\theta^*)\rangle$ for the ground state (obtained by a previous VQE iteration), we measure the 1-RDM $D_\mathrm{emb}$ of this state on the QPU, and we diagonalize it on a CPU to get the transformation $V$ to the NO basis corresponding to this guess, which is an operation polynomial in $N_c$. We then transform the Hamiltonian $H_\mathrm{emb}$ to this basis and perform a VQE in this basis (Fig.~\ref{fig:method}(c)), and so on until the energy converges.

Interestingly, this NOization amounts to yet another quantum-classical hybridization step, as the CPU is used to optimize the single-particle basis based on the QPU outputs. In a similar spirit - although more basic, an optimization of the ordering of the single-particle basis orbitals, based on a Mutual Information criterion, was proposed in Ref.~\onlinecite{tkachenko_correlation-informed_2021}.

\begin{figure}
    \centering
    \includegraphics[width=\columnwidth]{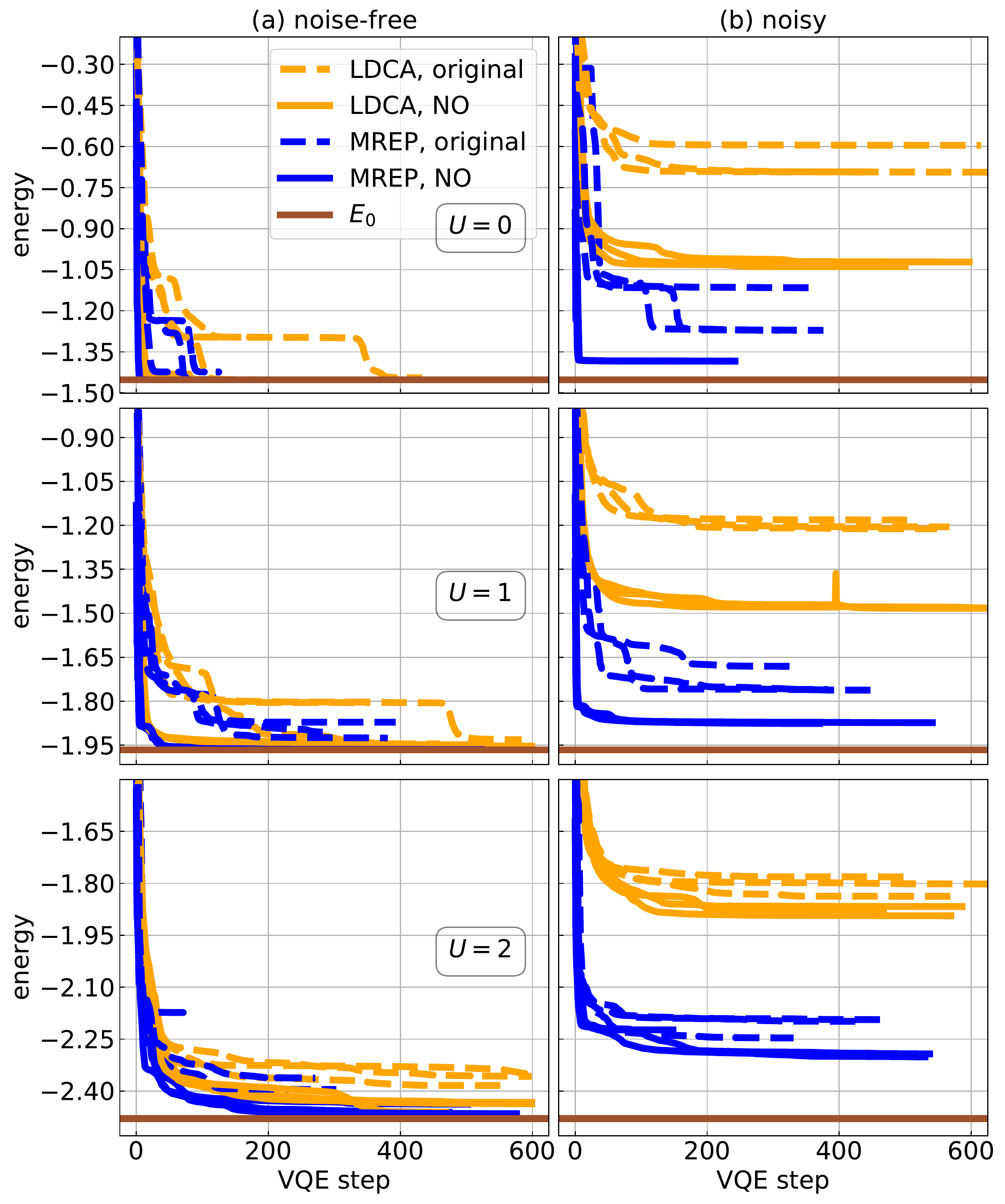}
    \caption{\textbf{Evolution of the variational energy as a function of the VQE step}, in the original vs Natural-Orbital (NO) basis, for the LDCA and MREP circuits,
    with three different (random) initializations of the circuit's parameters for each setup.
    {\it Left}: noiseless simulation. {\it Right}: noisy simulation.
    {\it Top to bottom}: $U=0, 1$ and $2$.
    $E_0$ denotes the exact GS energy of $H_\mathrm{emb}$. 
    }\label{fig:VQE}
\end{figure}

\begin{figure}
    \includegraphics[width=\columnwidth]{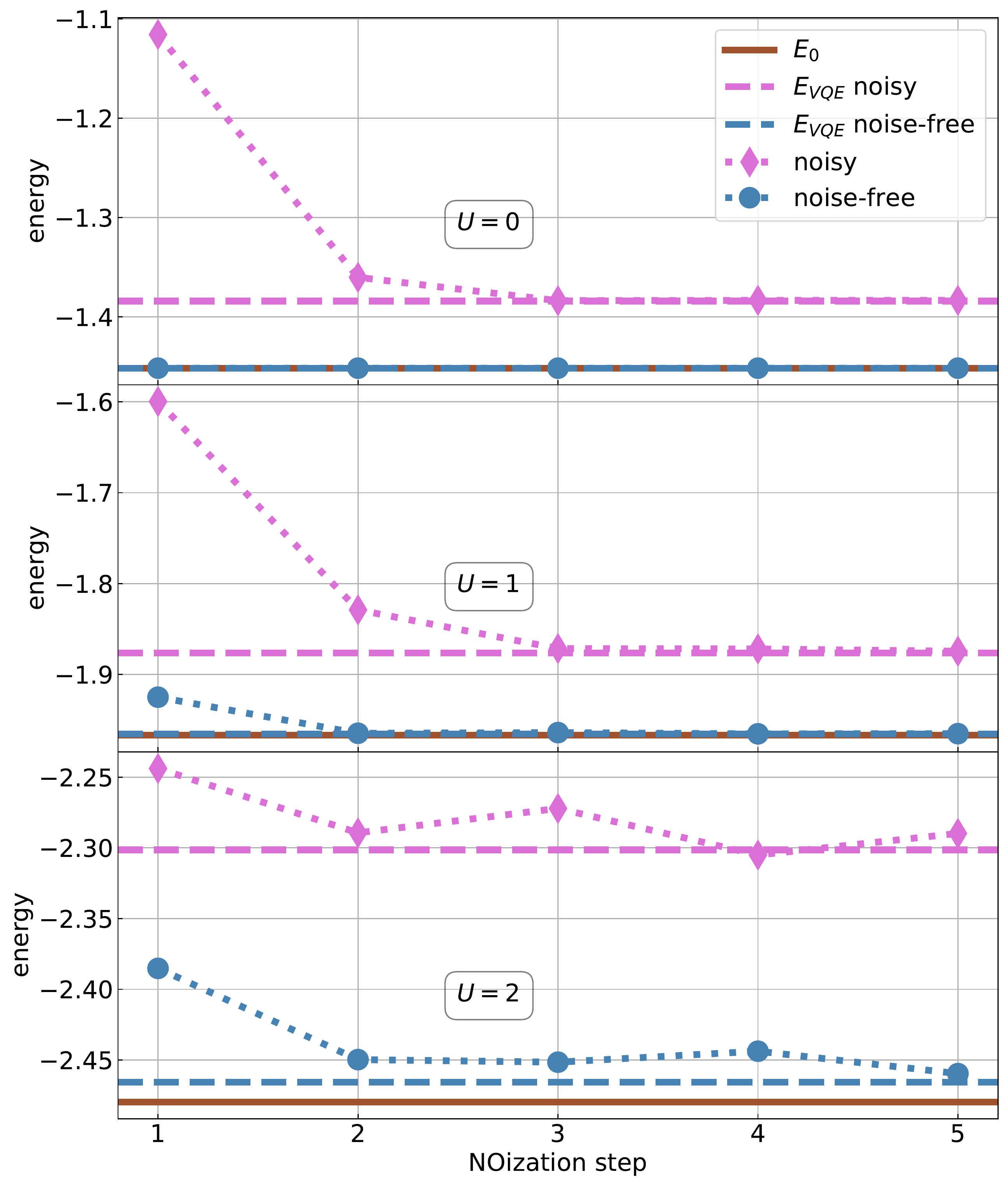}
    \caption{\textbf{Evolution of the minimal energy found by MREP-based VQE as a function of the NOization step}, with (dotted pink lines) or without (dotted blue lines) noise.
    {\it Top to bottom}: $U=0, 1$ and $2$.
    Dashed lines: minimal VQE energy found with {\it exact} NO basis.
    $E_0$ denotes the exact GS energy of $H_\mathrm{emb}$.
    }\label{fig:NOization}
\end{figure}

\begin{figure}
    \centering
    \includegraphics[width=\columnwidth]{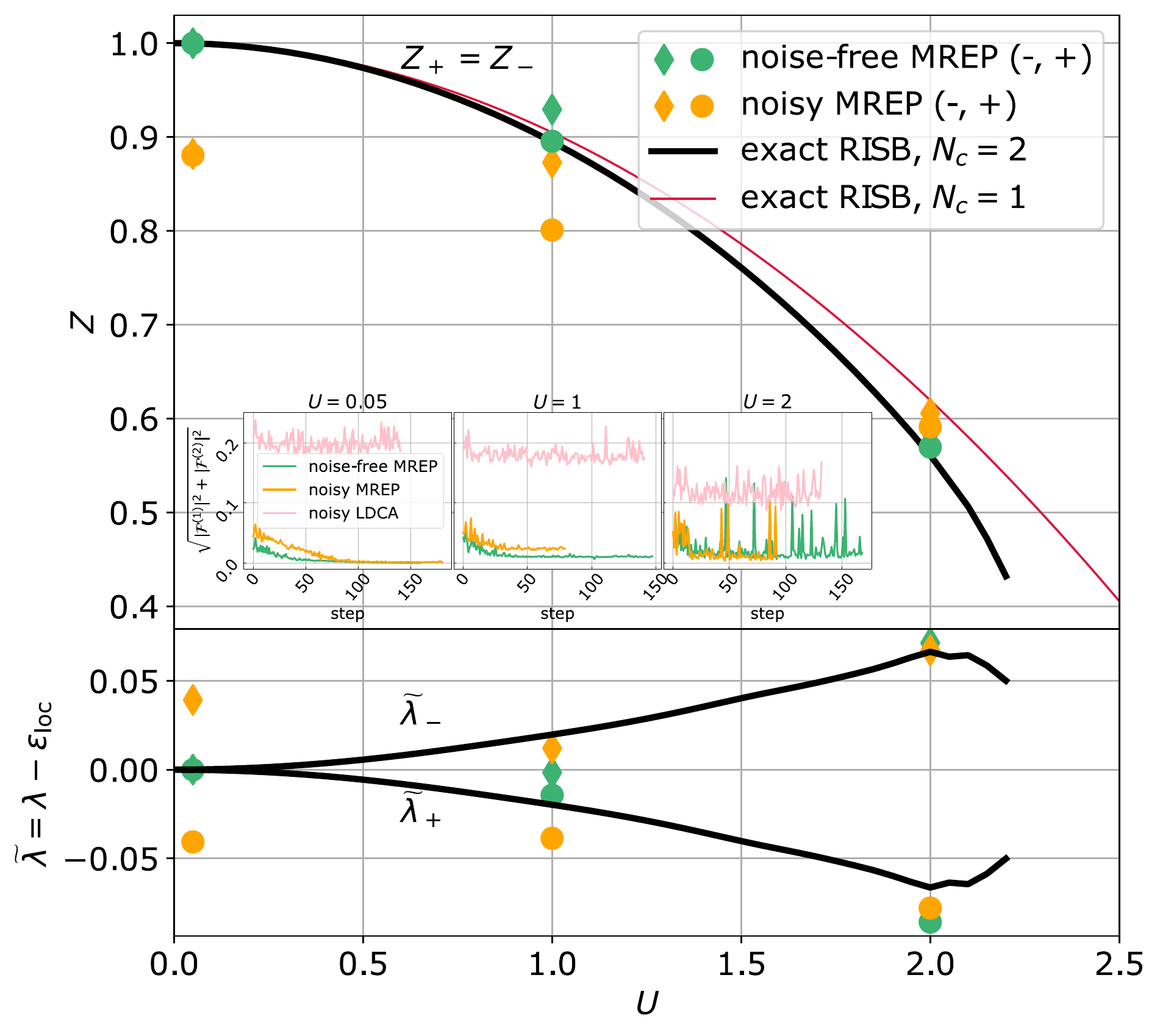}
    \caption{\textbf{Quasiparticle weight and static self-energy shift elements as a function of the on-site correlation.}
    Dots: RISB result using a noise-free or noisy QPU with the MREP ansatz.
    Solid lines: purely classical RISB results for $N_c=1,2$.
    Inset: evolution of the cost function along the $N_c=2$ RISB procedure for the MREP (with and without noise) and the noisy LDCA circuits. The "+" and "-" labels refer to the two independent matrix elements in the symmetry-adapted basis. Here, $\varepsilon_{\mathrm{loc}, -/+}=\mu \pm t$.
    }\label{fig:RISB_result}
\end{figure}

{\it Results.} We now turn to the results of this procedure for computing the properties of the Hubbard model using a $N_c=2$ impurity model within RISB embedding.
We first illustrate the power of NOization for both LDCA circuits and a new type of ansatz we dub the Multireference Excitation Preserving (MREP) ansatz, which bridges the gap between HEA and physically-motivated ansätze. Both ansätze are illustrated in Fig.~\ref{fig:circuits}. 
The MREP ansatz associates a quantum routine that prepares multireference states, which are key to quantum chemistry \cite{sugisaki_quantum_2019} and to the physics of impurity models \cite{snyman_efficient_2021},
with an excitation-number preserving routine that redistributes the fermionic excitations among the orbitals.
This latter routine contains an excitation-preserving two-qubit gate that is native to some superconducting processors (called \textit{fSim}\cite{kivlichan_quantum_2018} in the Sycamore processor\cite{arute_quantum_2019}, see also Refs.~\onlinecite{Barkoutsos2018, barron_preserving_2021}),
and that we use in a layered fashion.

In Fig.~\ref{fig:VQE}, we compare the results of VQE in the original site-spin basis and in the {\it exact} NO basis with and without noise. We choose a simple depolarizing noise model whose intensity is chosen to reproduce the gate error rates reported in NISQ processors (see Appendix \ref{sec:noise_models}), and we consider embedded Hamiltonians parametrized by ($R$, $\lambda$) converged with classical RISB at different levels of Coulomb interaction $U$.
We observe that working in the NO basis leads to faster convergence for both circuits, especially for the LDCA circuit for $U=0$ and $U=1$. VQE-converged NO states prepared with the MREP ansatz have a slightly lower energy than those in the original site-spin basis.
As noise is turned on, two phenomena occur: 
(i) the LDCA ansatz, due to its high depth (and hence, higher sensitivity to noise, as more errors are collected along the circuit), yields energies far off the mark, and 
(ii) the MREP ansatz yields significantly lower energies in the NO basis than in the original basis. 
The latter point substantiates the interest of working in the NO basis.
The efficiency of the MREP ansatz can be put in perspective with the recent observation \cite{bravyi_complexity_2017,debertolis_few-body_2021} that a rather modest number of Slater determinants should suffice to express the ground state of impurity models.

We now investigate the behavior of the iterative transformation to an approximate NO basis, since the exact NO basis is in principle unknown.
In Fig.~\ref{fig:NOization}, we show the evolution of the converged VQE energy with the NOization step for the MREP ansatz, with and without noise.
We observe that in all cases only a few (about three) steps lead to an energy very close to that attained by performing VQE directly in the exact NO basis, thus confirming the ability of the iterative procedure to provide a valid approximation of the NO basis.

We conclude with the results of the hybrid RISB approach in Natural Orbitals with and without noise, both for our MREP ansatz and for the LDCA circuit, with a full self-consistent convergence of the RISB external loop (see Appendix  \ref{sec:RISB} for details). 
Figure \ref{fig:RISB_result} shows the evolution of the quasiparticle weight and of the static self-energy with $U$, compared with the purely classical results for both $N_c=1$ and $N_c=2$. Each of those two quantities can be described by two matrix elements labeled "+" and "-" due to symmetry considerations (see Appendix \ref{subsec:symm} for further details).
We observe that the LDCA circuit, taken here as the reference to the prior state of the art, when run with a similar computational budget as the RISB with the MREP ansatz, yields very poor results in the presence of noise: we observe no convergence of the cost function (see inset of Fig.~\ref{fig:RISB_result}).
In contrast, the MREP ansatz used in the NO basis yields very accurate values for $Z$ and $\lambda$ in a noise-free setting, and remains quite accurate in the presence of realistic levels of noise.
The larger deviations at smaller $U$ values could be remedied by allowing for an adaptive construction of the ansatz \cite{grimsley_adaptive_2019}, which would enable the use of shorter circuits in the $U\rightarrow 0$ limit, where, in the NO basis, the ground state is weakly entangled.
Finally, the accuracy we reach allows us to resolve the coarse-grained space dependence of the static self-energy via its two components $\lambda_{\pm}$, a feature that was out of reach of previous ($N_c=1$) studies.

{\it Conclusion.}
In this work, we combined a novel iterative procedure on top of VQE, the NOization, which optimizes the single-particle basis representation of quantum circuits, with a variational ansatz circuit, the MREP ansatz, in order to variationally prepare the ground state of spatially extended embedded models with two correlated sites and two bath sites, as required by embedding methods for strongly correlated materials.
With the help of noisy simulations, we showed that this combination could handle NISQ-grade hardware noise and provide reasonably accurate and space-resolved ($N_c=2$) estimates of the quasi-particle renormalization weight and self-energy shift across a large range of interaction strengths in the Hubbard model. Furthermore, the NOization method allows for a straightforward resolution of the purely-local ($N_c=1$) problem examined in previous studies (see Appendix \ref{sec:Nc1}). 

Our work thus lays the groundwork for extending the size of the embedded problem, a crucial step for elucidating the physics of strongly correlated materials, where nonlocal fluctuations are key but can only be captured by  embedded models with $N_c > 1$.

Our work can be extended in several directions. While NOization addresses the most acute limitation of NISQ processors, namely their limited coherence, it may lengthen the duration for computing the expectation value of the Hamiltonian, as it usually increases the number of Pauli terms of the Hamiltonian. Whether this increase is compensated or not by a possible decrease in the individual statistical error (because of a more peaked distribution) remains to be investigated.
Similarly, the question of the robustness of the classical VQE optimization to shot noise is an important one. Shot-noise-robust approaches \cite{nakanishi_sequential_2020,ostaszewski_structure_2021} necessitating only three energy evaluations (via the parameter-shift rule), but hitherto limited to one-qubit parameterized gates, were recently extended to parameterized gates acting on more qubits \cite{izmaylov_analytic_2021}, which should provide a similar shot-noise-resilient optimization scheme.

\textit{Note added.} While preparing this manuscript, a more advantageous circuit \cite{steckmann_simulating_2021} than the one \cite{keen_quantum-classical_2020} we used as prior art reference to treat the one-impurity case in Appendix \ref{sec:Nc1} was proposed.  Although it has a slightly larger gate count than the multireference circuit we use in our proposal, it is also single-parameter and works in the original basis.

\begin{acknowledgments}
We thank Michel Ferrero, Simon Martiel as well as Tsung-Han Lee for many useful discussions. The computations were performed on the Atos Quantum Learning Machine (QLM).
\end{acknowledgments}

\appendix

\renewcommand\thefigure{A\arabic{figure}}
\setcounter{figure}{0}

\section{Quantum computation details}

\subsection{Encoding}
To go from fermionic variables to spin variables to write the Hamiltonian, one must provide an encoding scheme. Here we use the straightforward Jordan-Wigner encoding, mapping the creation/annihilation operators as:
\begin{align}
    & c^{\dagger}_j \rightarrow \hat{Z}_1 \otimes \hat{Z}_2 \otimes ... \otimes \hat{Z}_{j-1} \otimes \frac{1}{2} (\hat{X}_j + i \hat{Y}_j) \\
    & c_j \rightarrow \hat{Z}_1 \otimes \hat{Z}_2 \otimes ... \otimes \hat{Z}_{j-1} \otimes \frac{1}{2} (\hat{X}_j - i \hat{Y}_j)
\end{align}
where the chains of $\hat{Z}$ operators ensure fermionic anticommutation.

\subsection{Ansatz circuits}

\subsubsection{Low-Depth Circuit Ansatz (LDCA)}
\begin{figure}
    \centering
    \includegraphics[width=0.8\columnwidth]{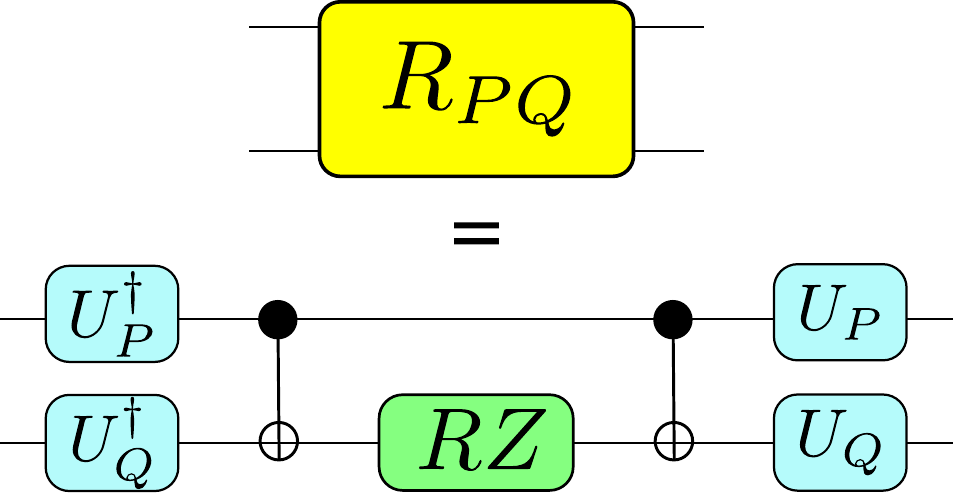}
    \caption{Decomposition of a 2-qubit rotation gate. Here, $P, Q \in X, Y, Z$ are rotation axes and $U_X=RY(\frac{\pi}{2}), U_Y=RZ(\frac{\pi}{2}) RY(\frac{\pi}{2})$ and $U_Z=I$.
    }
    \label{fig:RPQ_gate}
\end{figure}

The Low-Depth Circuit Ansatz\cite{dallaire-demers_low-depth_2019}  (exemplified in its eight-qubit, one-cycle version on Figure \ref{fig:circuits}) is inpired from a special class of circuits 
meant to prepare uncorrelated (or "gaussian") states (see Refs \onlinecite{jozsa_matchgates_2008} and  \onlinecite{dallaire-demers_application_2020}).
The building block of uncorrelated states preparation is the sequence of two-qubit rotations $R_{YY}(\theta_1) R_{XX}(\theta_2) R_{YX}(\theta_3) R_{XY}(\theta_4)$, with
\begin{equation}
    R_{PQ}^{(i,j)}(\theta) = e^{i \theta P_i \otimes Q_j}
\end{equation}
arranged in so-called matchgate cycles that alternatively connect qubits $i, i+1$ and $i+1, i+2$.
To endow the output state with a non-gaussian (i.e correlated) character, a $R_{ZZ}$ rotation is inserted in the $YY$-$XX$-$YX$-$XY$ sequence.

Decomposing the two-qubit gates into CNOT and one-qubit rotation gates (see Figure \ref{fig:RPQ_gate}), the one-cycle version of the LDCA circuit with eight qubits we have used to prepare our embedded models' ground states has initially a total count of 1108 gates, including 280 CNOT gates. Each matchgate sequence of five two-qubit rotation gates comprises twenty-five one-qubit gates (up to the cancellation due to the successive application of $U_P$ and $U_P^{\dagger}$) and ten two-qubit gates. Yet, it was shown that a two-qubit unitary could be written with at most fifteen one-qubit gates and three CNOT gates \cite{vatan_optimal_2004}. We thus applied circuit recompilation techniques \cite{martiel_architecture_2020} to lower this count and achieved a count of 324 gates, among which 112 are CNOT gates.

\subsubsection{Multireference Excitation Preserving (MREP) circuit}

The Multireference Excitation Preserving (MREP) ansatz is a custom-made ansatz that starts from a multireference state and then distributes the fermionic excitations among the orbitals.

We design the sequence of the first few gates (one-qubit gates and subsequent CNOT gates) by duplicating the gate patterns that we use for the single-site case (see Appendix~\ref{sec:Nc1} below), which are themselves inspired by Ref.~\onlinecite{sugisaki_quantum_2019}.
This effectively corresponds to starting with a circuit that would capture exactly the ground state of two disconnected (without intra-dimer hopping) single-site impurity problems.
 
The second part of the MREP circuit is composed of layers of $U_F$ blocks, each of which contains cone-shaped patterns of so-called \textit{fSim} gates\cite{kivlichan_quantum_2018, Foxen2020}, defined as\cite{arute_quantum_2019}:
\begin{equation}
    fSim(\theta,\phi)=\left[\begin{array}{cccc}
1 & 0 & 0 & 0\\
0 & \cos\theta & -i\sin\theta & 0\\
0 & -i\sin\theta & \cos\theta & 0\\
0 & 0 & 0 & e^{-i\phi}
\end{array}\right].
\end{equation}
These gates are native to transmon-qubit architectures such as the Sycamore chip. 
From the point of view of fermionic physics, \textit{fSim} gates can create gaussian (i.e uncorrelated) states when $\phi = 0$ (they then belong to the set of so-called "matchgates" that are universal for simulating uncorrelated fermions).
Conversely, with $\phi \neq 2n\pi $ ($n\in \mathbb{Z}$), they create states with a non-gaussian character.
Importantly, they do not modify the number of fermions in the wavefunction, which is why they are described as "excitation preserving". 
Thus, the \textit{fSim} "cones" allow for the spreading of the excitations among the orbitals. This is somewhat akin to turning on the intradimer hopping.

In the main text, we chose a number of four layers of \textit{fSim} gates. This choice is dictated by the numerical observation that more layers induced only a marginal improvement of the VQE energies.

\subsection{Classical optimization} \label{subsec:classical_optim}
In this subsection, we focus on the classical optimizer used within the VQE procedure described in Fig.~\ref{fig:method}(c).

The circuits are optimized using Python's scipy optimization package with the BFGS method, allowing for up to 10000 iterations of the algorithm. 

To account for the dependence of the accuracy of the VQE optimization to the initialization of the circuit's parameters, three runs were made for each configuration, and one given random set of parameters was used as a first guess twice: for a subsequent noise-free optimization as well as a noisy one.

\section{Noise models}\label{sec:noise_models}

In this appendix, we describe the noise models used to simulate noisy QPUs.

NISQ processors experience many kinds of errors, among which the most prominent ones are gate noise, idling noise and State Preparation and Measurement (SPAM) errors.
In this work, we deliberately choose a simple noise model with only gate noise and adjust the level of this gate noise to match the error levels measured in Randomized Benchmarking experiments on current NISQ processors.

More specifically, we work with a simple depolarizing noise model: the density matrix $\rho$ evolves after a one-qubit process according to the channel
\begin{equation}
    \mathcal{E}^{(1)}(\rho) = (1-p_1)\rho + \frac{p_1}{3}\left( X \rho X + Y \rho Y + Z \rho Z \right),
\end{equation}
which can be interpreted as a random Pauli operation being inserted with depolarization probability $p_1$, and the expected gate operation occurring alone with probability $(1-p_1)$.
Likewise, two-qubit gate errors are modelled as a two-qubit depolarizing channel that consists of the tensor product of the 1-qubit channels with depolarizing probability $p_2$.

The $p_1$ and $p_2$ probabilities are taken to reproduce randomized-benchmarking error levels.
Starting from the relationship between average process fidelity and the coefficient $p_0$ of the identity in the Kraus decomposition of the channel \cite{magesan_characterizing_2012},
\begin{equation}
    \mathcal{F}_{\mathrm{ave}} \equiv 1-\epsilon_\mathrm{RB} = \frac{p_0d+1}{d+1},
\end{equation}
with $d=2^n$ the dimension of the subspace that is acted on by the channel, we must set $p_0$ so that
\begin{equation}
    p_0 = 1-\left(1+\frac{1}{d} \right)\epsilon_\mathrm{RB}.
\end{equation}

Applying this formula to the one-qubit process, we get
\begin{equation}
    p_1 = \frac{3}{2} \epsilon_\mathrm{RB}^{(1)},
\end{equation}
and for the two-qubit process:
\begin{equation}
    p_2 = 1-\sqrt{1-\frac{5}{4}\epsilon_\mathrm{RB}^{(2)}}.
\end{equation}

The specific values we choose for $\epsilon_\mathrm{RB}^{(1)}$ and $\epsilon_\mathrm{RB}^{(2)}$ are those measured through randomized benchmarking performed on Google's Sycamore chip (see the supplemental material of Ref.~\onlinecite{arute_quantum_2019}):
\begin{align}
    & \epsilon_\mathrm{RB}^{(1)} = 0.16\%,\\
    & \epsilon_\mathrm{RB}^{(2)} = 0.6 \%.
\end{align}

\section{The Rotationally Invariant Slave Boson method}\label{sec:RISB}
In this appendix, we elaborate on the embedding method used to solve the Hubbard model in the main text.

\subsection{Formalism}

There are \textit{a priori} several embedding methods that could be used to map the original lattice problem, the Hubbard model, onto an effective problem of reduced dimensionality. Prominent choices include, ranging from the most accurate (and computationally expensive) to the simplest, Dynamical Mean Field Theory (DMFT), Rotationally-Invariant Slave Boson (RISB, which is equivalent to the Gutzwiller method used in Ref.~\onlinecite{yao_gutzwiller_2021}), and Density-Matrix Embedding Theory (DMET). DMET can essentially be regarded as a simplified version of RISB, which itself can be regarded as a low-energy version of DMFT \cite{ayral_dynamical_2017}.
Here, as briefly argued in the main text, we choose to resort to RISB: on the one hand, as a low-energy simplification of DMFT, it does not require the computation of the full time-dependent Green's function, nor the use of a large number of uncorrelated bath orbitals, both of which would be costly in terms of the depth and width (respectively) of the ansatz quantum circuits required by the VQE step.
On the other hand, RISB, contrary to DMET, gives access to the quasiparticle renormalization factor $Z$, a key quantity to get insights into the properties of correlated materials. 
We note that this approach shares commonalities with "Energy-Weighted DMET"\cite{Fertitta2018, Fertitta2019}, which interpolates between DMET and DMFT by requiring an increasing number of bath levels to progressively describe the full dynamics of the self-energy. This approach was implemented for the $N_c=1$ case in Ref.~\onlinecite{tilly_reduced_2021}.

We use the RISB method as introduced by Ref.~\onlinecite{lechermann_rotationally-invariant_2007} as a rotationally invariant generalization of the work by Kotliar and Ruckenstein \cite{kotliar_new_1986}. It allows to properly handle multi-orbital problems, whether the orbitals denote true atomic orbitals or, as in the current work, the $2 N_c$ site orbitals of a unit cell (counting spin degeneracy). 
We use a recent refinement of RISB \cite{lanata_phase_2015} where the free energy functional---\textit{a priori} a highly nonlinear function of the slave-boson amplitudes $\Phi_{An}$---is made quadratic in the amplitudes, at the expense of adding Lagrange multipliers and additional variables.
The resulting six-variable free energy is extremized to find the six Lagrange equations of the problem:
\begin{subequations}
\begin{align}
 & \Delta_{\alpha\beta}^{p}=\sum_{\boldsymbol{k}\in\mathrm{BZ},i\omega}\left[i\omega-R \varepsilon_{\boldsymbol{k}}R^{\dagger}-\lambda+\mu\right]_{\beta\alpha}^{
-1}e^{i\omega0^{+}},\label{eq:Delta_p-lagrange}\\
 & \sum_{\mu}\left[\left(\Delta^{p}(1-\Delta^{p})\right)^{1/2}\right]_{\alpha\mu}\mathcal{D}_{\beta\mu}\label{eq:D_lagrange}\\
 & \;=\sum_{\boldsymbol{k}\in\mathrm{BZ},i\omega}\left[\left\{ \varepsilon_{\boldsymbol{k}}R^{\dagger}\right\} \left[i\omega-R\varepsilon_{\boldsymbol{k
}}R^{\dagger}-\lambda+\mu\right]^{-1}\right]_{\beta\alpha}e^{i\omega0^{+}},\nonumber \\
 & \lambda_{\alpha\beta}^{c}=-\lambda_{\alpha\beta}\nonumber\\
 &\;-\sum_{\gamma\delta\eta}\left\{ \mathcal{D}_{\gamma\delta}R_{\eta\gamma}\frac{\partial\left[\left(\Delta^{p}(1-\Delta^{p})\right)^{1/2}\right]_{\eta\delta}}{\partial\Delta_{\alpha\beta}^{p}}+c.c\right\} ,\label{eq:lambda_c-lagrange}\\
 & \hat{H}_{\mathrm{emb}}|\Phi\rangle=E_0|\Phi\rangle,\label{eq:H_imp-1}\\
 & \langle\Phi|f_{\beta}f_{\alpha}^{\dagger}|\Phi\rangle=\Delta_{\alpha\beta}^{p},\label{eq:F1}\\
 & \langle\Phi|c_{\alpha}^{\dagger}f_{\beta}|\Phi\rangle=R_{\gamma\alpha}\left[\left(\Delta^{p}(1-\Delta^{p})\right)^{1/2}\right]_{\gamma\beta}.\label{eq:F2}
\end{align}
\end{subequations}
Here, $\varepsilon_{\boldsymbol{k}}$ is the free dispersion on a square lattice tiled with $N_c=2$-site unit cells (as in, e.g., Ref.~\onlinecite{lee_rotationally_2019}). $i\omega$ denotes Matsubara frequencies, and $\boldsymbol{k}\in \mathrm{BZ}$ denotes discretized points in the first Brillouin zone.

$H_\mathrm{emb}$ is the impurity model defined in Eq.~\eqref{eq:Hemb} of the main text, with $2 N_c$ correlated orbitals ("impurities") and $2 N_c$ bath orbitals.
Technically, the appearance of an impurity model in RISB comes from a reinterpretation of the slave-boson amplitudes $\Phi_{An}$ as the coefficients of the Schmidt decomposition of a ket $|\Phi\rangle$ defined on a Hilbert space that is a tensor product of the original $2 N_c$ fermionic degrees of freedom with an additional $2 N_c$ bath degrees of freedom\cite{lanata_phase_2015}.
The Greek indices are compound indices $\alpha = (i, \sigma)$ with $i=1\dots N_c$ and $\sigma=\uparrow,\downarrow$.
We refer the reader to Ref.~\onlinecite{ayral_dynamical_2017} for a derivation of these equations and an explanation of the meaning of the $\Delta^p$, $\mathcal{D}$, $\lambda^c$ variables.
As explained in the main text, $R$ and $\lambda$ turn out to be low-energy parametrizations of the lattice self-energy via Eq.~\eqref{eq:selfenergy}.

To solve these equations, we reformulate them as a root problem as in e.g Ref.~\onlinecite{ayral_dynamical_2017}: we seek to find the roots $R$ and $\lambda$ of the functions
\begin{subequations}
\begin{align}
\mathcal{F}^{(1)}[R,\lambda] & \equiv\langle\Phi|f_{\beta}f_{\alpha}^{\dagger}|\Phi\rangle-\Delta_{\alpha\beta}^{p}\label{eq:F1_def},\\
\mathcal{F}^{(2)}[R,\lambda] & \equiv\langle\Phi|c_{\alpha}^{\dagger}f_{\beta}|\Phi\rangle-R_{\gamma\alpha}\left[\left(\Delta^{p}(1-\Delta^{p})\right)^{1/2}\right]_{\gamma\beta},\label{eq:F2_def}
\end{align}
\end{subequations}
where $\mathcal{F}^{(1)}$ and $\mathcal{F}^{(2)}$ are implicit functions
of $R$ and $\lambda$ via the above equations.

The computational bottleneck of the root-solving procedure is the solution of the impurity model, Eq.~\eqref{eq:H_imp-1} that is needed to compute the 1-RDM elements $\langle\Phi|f_{\beta}f_{\alpha}^{\dagger}|\Phi\rangle$ and $\langle\Phi|c_{\alpha}^{\dagger}f_{\beta}|\Phi\rangle$ and then $\mathcal{F}^{(1)}$ and $\mathcal{F}^{(2)}$.
While it is usually solved with a classical impurity solver (by, e.g., exact diagonalization of $H_\mathrm{emb}$), we propose to solve it approximately using the hybrid quantum-classical VQE method combined with the NOization procedure described in the Appendix \ref{sec:NOization}.

\subsection{Implementation details}{\label{subsec:symm}}

For the solution of the RISB equations, we discretize the first Brillouin zone with a regular two-dimensional mesh with $32 \times 32$ points. We use an inverse temperature of $\beta = 300$ (instead of $\beta = \infty$, in order to smoothen the summation). We take $t = -0.25$, so that the half-bandwidth is unity.

While the $R$, $\lambda$, $\Delta_p$, $\mathcal{D}$, and $\lambda^c$ variables are in principle $2 N_c \times 2 N_c$ matrices, the absence of spin-flip terms in the Hamiltonian and the enforcement of paramagnetism allows to simplify all those matrices as $A_{i\sigma, j \sigma'} = \tilde{A}_{ij} \delta_{\sigma \sigma'}$.
Finally, the mirror symmetry of the $N_c=2$ unit cell obtained by periodizing the lattice implies, in the absence of symmetry  breaking, that in the symmetry-adapted basis one has $\tilde{A}_{ij} = 
\left[\begin{array}{cc}
a & b\\
b & a
\end{array}\right]
$ so that for each of these matrices has only two degrees of freedom.
We use these symmetry properties to simplify the computations. We refer the reader to Refs~\onlinecite{lanata_efficient_2012, lanata_phase_2015} for more details on these symmetry considerations.

\subsection{Convergence details}

\begin{figure}
    \centering
    \includegraphics[width=\columnwidth]{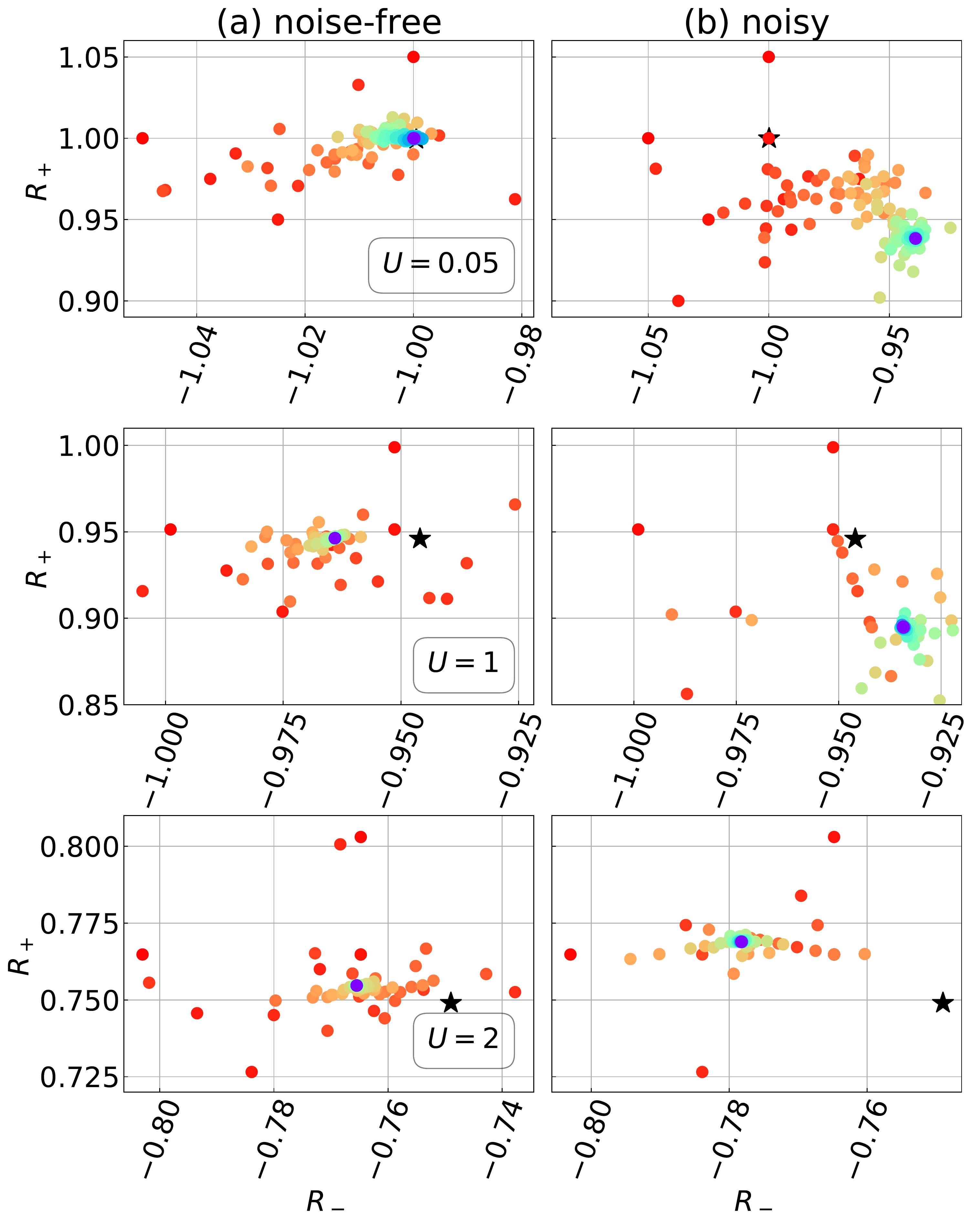}
    \caption{Trajectory of the two components $R_+$ and $R_-$ of $R$ along the $N_c=2$ RISB minimization procedure based on the VQE optimization of the MREP ansatz in Natural Orbitals, with (right column) and without (left column) noise, for $U=0.05, 1, 2$. The black star materializes the classical RISB solution. The colormap enables to visualize the progression from the first guess (deepest red shade) to the last iteration (in purple).}
    \label{fig:R}
\end{figure}

While the solution of RISB equations can be formulated as a root-solving problem, we found that the use of an approximate impurity solver caused the root-solving procedure to fail.
We therefore turned the root-solving procedure into a minimization problem, in which $\sqrt{||\mathcal{F}^{(1)}||^2+||\mathcal{F}^{(2)}||^2}$(in which $\mathcal{F}^{(1/2)}$ matrices were turned into vectors) is minimized.
We used the Nelder-Mead algorithm to solve this minimization problem.

We allow for up to 100 iterations of the algorithm, taking the classical solution $(R, \lambda)$ at $U-0.05$ as a starting point.

To ensure we indeed get a minimization, we plotted the evolution of the cost function as evaluated along the minimization procedure (see the inset in Fig.~\ref{fig:RISB_result}, where it was considered that the convergence had been reached even though the whole computational budget was not necessarily consumed). In the absence of noise, the MREP ansatz-based does provide a clear RISB minimization of the cost function for $U=0.05$ and $U=1$. For $U=2$, after a regime in which the cost function diminishes in average, there is a plateau regime affected by spikes where the cost function gets very high. We checked that these spikes were caused by sensitively larger VQE errors at these points. The convergence is still manifest in the presence of noise, although less pronounced. On the other hand, the LDCA ansatz gives rise to a noisy RISB loop that does not converge. We conclude that the latter ansatz is ruled out for RISB at cluster size $N_c=2$ with current noise levels.
The trajectory of the components of $R$ along the minimization is also presented, on Fig.\ref{fig:R}. We see that one of the components accurately converges whereas the other one gets deviated.

Note that to accelerate the computation, we carried out the RISB procedure directly in the exact NO basis, obtained by diagonalizing the Hamiltonian. 
We checked that the NOization procedure does not impede the RISB convergence by running the optimization for $U=1$ with the NOization layer (see Fig.~\ref{fig:conv_NO_VS_NOization}). 

The VQE output state for which a 1-RDM is computed and diagonalized is the lowest-energy lying state obtained by running five VQE optimizations corresponding to five different, random initializations for the MREP ansatz. For the LDCA ansatz, only one run is considered to limit the computational overhead. This is not expected to impede the procedure since there is little dependence to the initial condition according to Fig.~\ref{fig:VQE}(b).

Fig.~\ref{fig:RISB_result} display results of the RISB procedure in terms of quasi-particle weights $Z_{\pm} = |R_{\pm}|^2$ and non-trivial static self-energy shift elements $\widetilde{\lambda}_{\pm} = \lambda_{\pm} - \varepsilon_{\mathrm{loc}}$ (where $\varepsilon_{\mathrm{loc}} \equiv \sum \limits_{\boldsymbol{k} \in BZ}\varepsilon_{\boldsymbol{k}} - \mu$).

\section{Natural Orbitalization}\label{sec:NOization}

In this section, we give more details about the Natural Orbitalization (NOization) method introduced in the main text.

\subsection{Formalism}

Our goal is to express our problem in the basis of the Natural Orbitals. However, this basis can only be computed once the exact ground state---which we are looking for---is known. 
It is thus not directly accessible.
Nonetheless, it is possible to approximately rotate into this basis. We do this by applying the VQE procedure several times, using the variational approximation to the ground state provided by the VQE algorithm at step $k$ to update the orbital basis in which VQE at step $k+1$ is run.

More specifically, provided the optimal VQE state $\ket{\psi(\vec{\theta}^{*(k)})}$ returned by the VQE procedure, we use the quantum computer to compute the 1-RDM 
\begin{equation}
    D_\mathrm{emb} [\psi(\vec{\theta}^{*(k)}), c^{(k) \dagger}, c^{(k)}] \equiv \langle \psi(\vec{\theta}^{*(k)}) | c^{(k) \dagger}_i c^{(k)}_j |\psi(\vec{\theta}^{*(k)}) \rangle,
\end{equation}
and classically compute the transformation $V^{(k)}$ that diagonalizes it:
\begin{equation}
D_\mathrm{emb} [\psi(\vec{\theta}^{*(k)}), c^{(k) \dagger}, c^{(k)}]_{pq} = V^{(k)}_{p\alpha} n_{\alpha} V^{(k)\dagger}_{\alpha q}.
\end{equation}
where the Einstein convention is used on repeated indices.\\
This matrix is used to update the orbital basis as:
\begin{equation}
    c^{(k+1)\dagger}_{\alpha} = \sum_p V^{(k)}_{p\alpha} c^{(k)\dagger}_p,
\end{equation}
which corresponds to the following transformation on the Hamiltonian's coefficients:
\begin{align}
    &h^{(k+1)}_{pq}=V^{(k)}_{p'p}h^{(k)}_{p'q'}(V^{(k)\dagger})_{qq'} \\
    &h^{(k+1)}_{pqrs}=V^{(k)}_{p'p}V^{(k)}_{q'q}h^{(k)}_{p'q'r's'}(V^{(k)\dagger})_{rr'}(V^{(k)\dagger})_{ss'}.
\end{align}
Note that since in general the accuracy of the VQE procedure is sensitive to the initial tuning of the parameters, we may want to consider the best VQE run out of several ones corresponding to different initializations. The plots of Fig.~\ref{fig:NOization} were however obtained by running one single VQE from a random initialization for each point, and may thus be slightly impacted by this effect.

\subsection{RISB convergence and NOization: data}

\begin{figure}
    \centering
    \includegraphics[width=0.8\columnwidth]{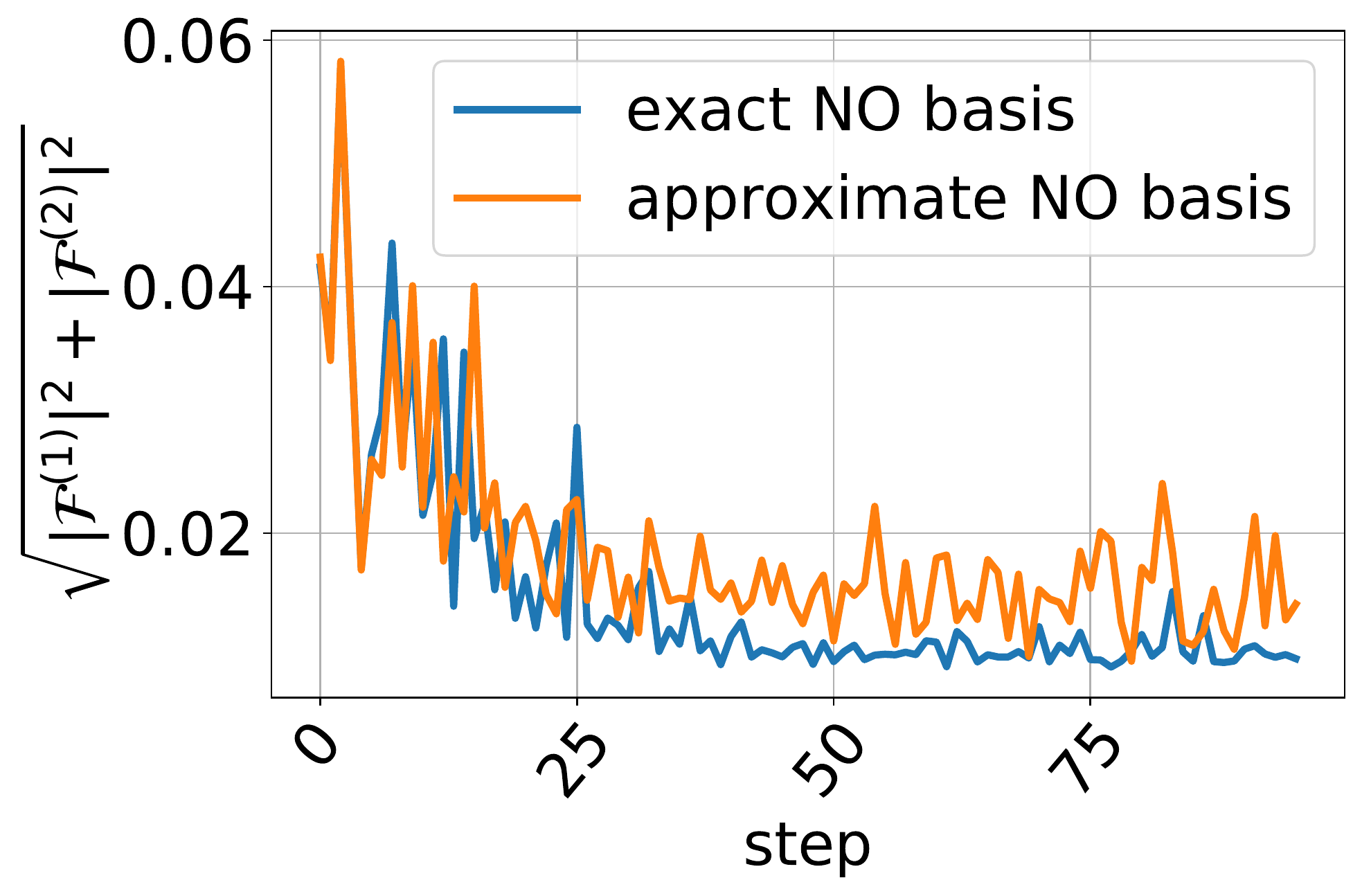}
    \caption{Convergence of the cost function for $N_c=2$ RISB at $U=1$ with the MREP ansatz without noise, in the exact NO basis (blue line) and in an effective NOization setup in which the NO basis is approximated iteratively (orange line).}
    \label{fig:conv_NO_VS_NOization}
\end{figure}

A display of RISB convergence with the effective NOization procedure against the RISB convergence exhibited running VQE in the exact NO basis is presented in Fig.~\ref{fig:conv_NO_VS_NOization} for $U=1$, showing that the procedure does indeed work in the context of RISB minimization (although the cost function converges to a slightly higher value).

\section{Single-site results}\label{sec:Nc1}

\begin{figure}
    \centering
    \includegraphics[width=0.8\columnwidth]{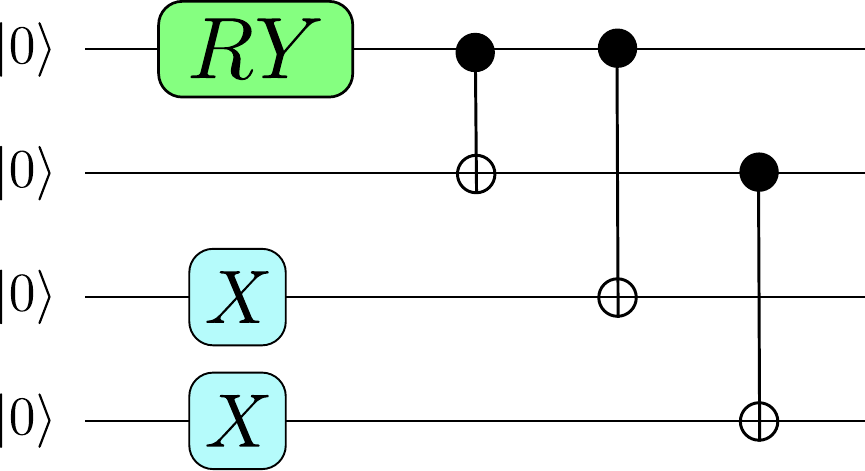}
    \caption{The Multireference (MR) state preparation circuit ($N_c=1$ case).}
    \label{fig:MR_circuit}
\end{figure}

\begin{figure}
    \centering
    \includegraphics[width=\columnwidth]{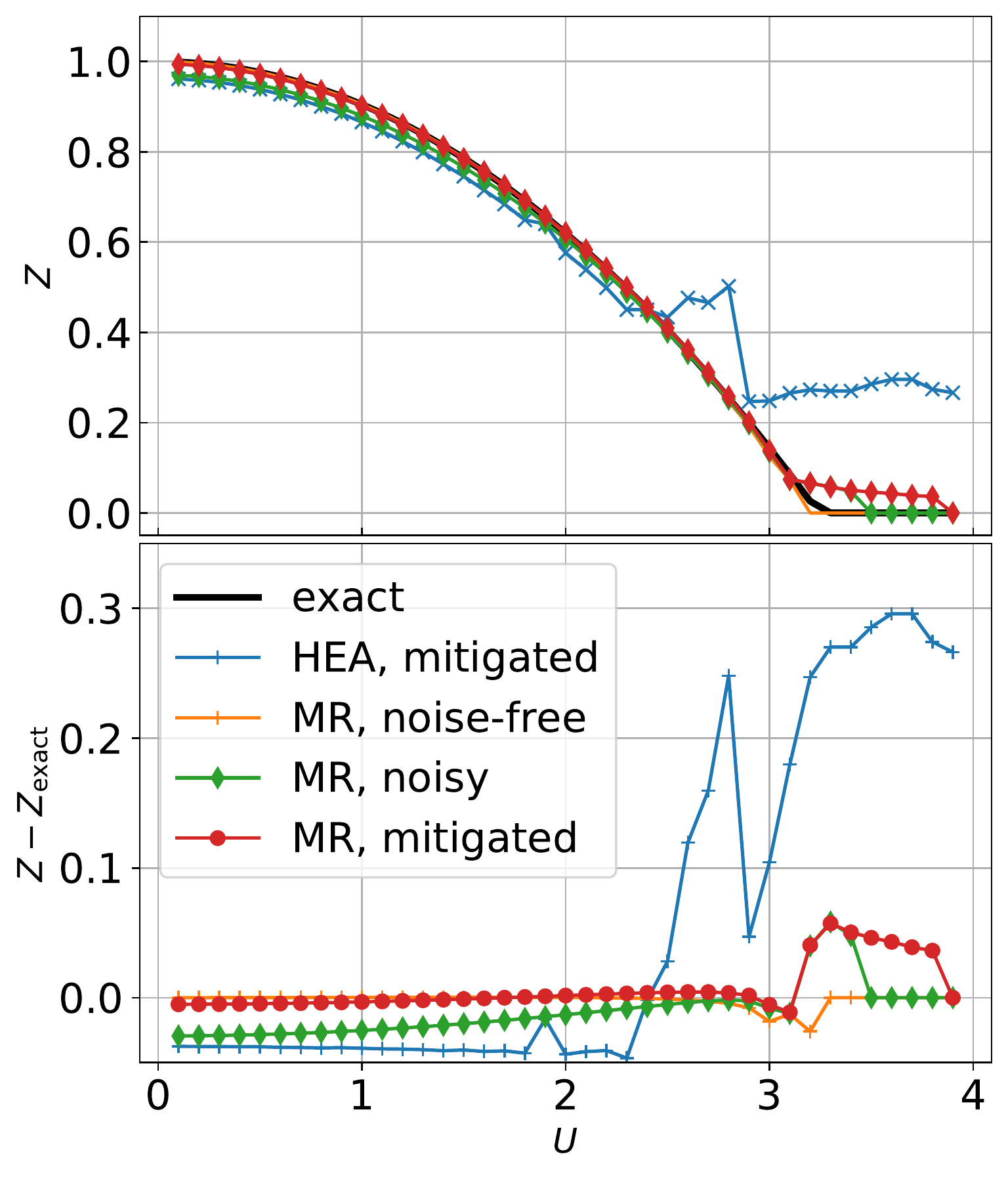}
    \caption{Evolution of the quasiparticle weight $Z$ as a function of $U$ for $N_c=1$ (top) and distance to the analytical solution (bottom) for the Multireference ansatz (MR, this work) compared with the HEA ansatz.
    }
    \label{fig:Z_Nc1}
\end{figure}

\begin{figure}
    \centering
    \includegraphics[width=\columnwidth]{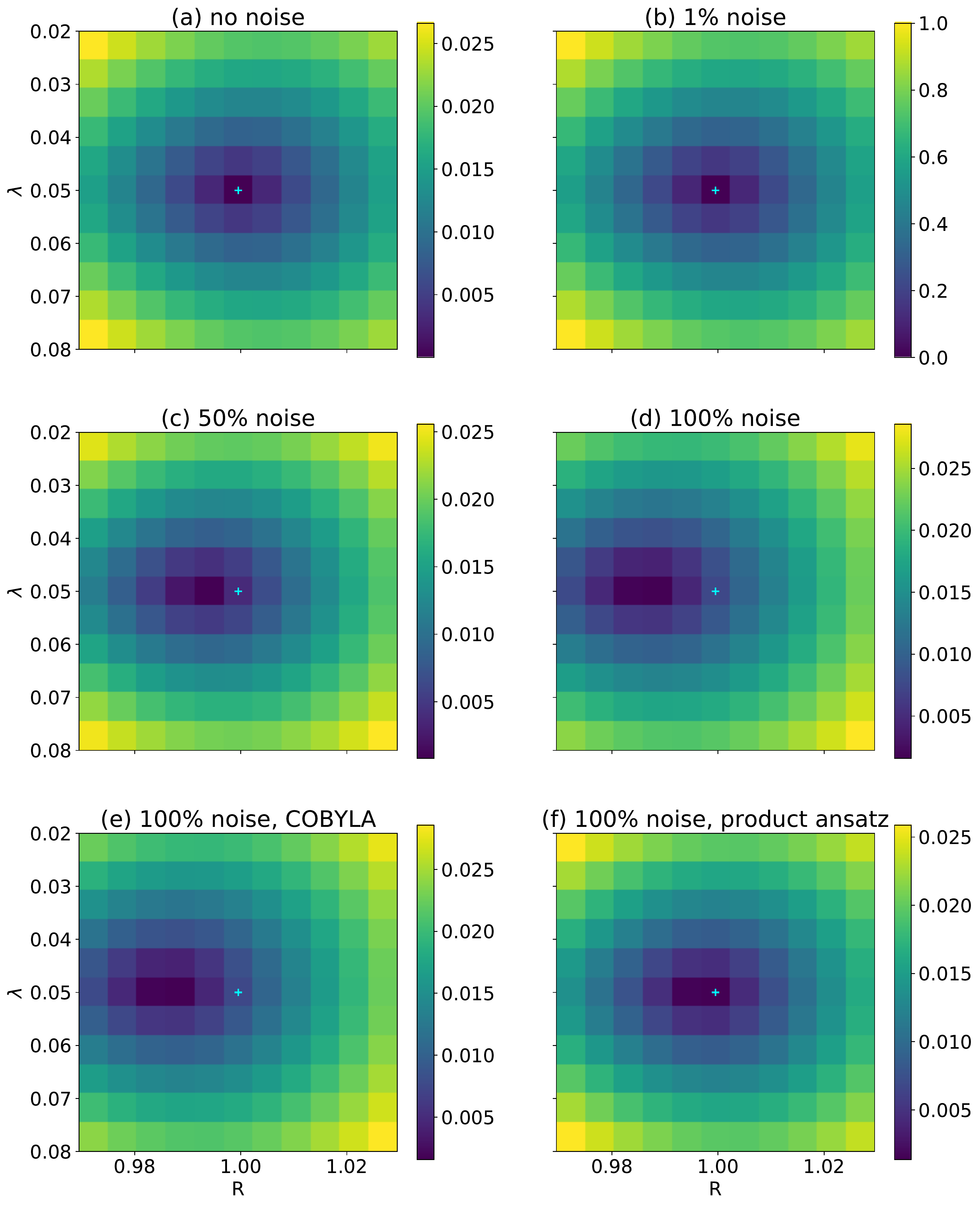}
    \caption{Optimization landscape $(R, \lambda$) around the analytical solution for $U=0.1$, $N_c=1$. The blue cross materializes the solution. From (a) to (d), each point is computed using the parameter shift rule to  tune the MR circuit with depolarizing noise levels set at an increasing fraction of the Sycamore chip's levels. (e): landscape associated with a full noise MR circuit optimization made with the COBYLA algorithm. (f): landscape obtained using a $RY$ gates-product ansatz, optimized with COBYLA with full noise on.}
    \label{fig:optim_ls_Nc1}
\end{figure}

While the main text focuses on impurity models with $N_c=2$ correlated sites, which in principle allows to obtain a coarse-grained space dependence of the correlation parameters $R$ and $\lambda$, it is also instructive to look at the single-site case ($N_c=1$), where the impurity model is said to be purely local and $R$ and $\lambda$ are merely scalar numbers. This case was addressed in a quantum-classical fashion in Refs~\onlinecite{keen_quantum-classical_2020,rungger_dynamical_2020,yao_gutzwiller_2021,tilly_reduced_2021}. We recall that we are still solving for the properties of a large or infinite-dimensional Hubbard model defined in Eq.~\eqref{eq:Hubbard}, a large $N_c$ giving information on more and more extended correlation effects. We refer the reader to Ref.~\onlinecite{lee_rotationally_2019} for a study of the $N_c$-dependence of RISB.

\subsection{Circuit}
For the paramagnetic RISB embedding model at $N_c=1$ at half-filling, one can write an exact ansatz circuit to prepare the natural-orbital ground state. Indeed, it is easy to show that in this case, the "natural ground state"'s 1-RDM is of the form
\begin{equation}
\label{eq:RDM_NO}
    D_\mathrm{emb} = \begin{pmatrix}
    n_0 & 0& 0& 0\\
    0& n_0& 0& 0\\
    0& 0& 1-n_0& 0\\
    0& 0& 0& 1-n_0
    \end{pmatrix}
\end{equation}
The 1-RDM in this case characterizes uniquely the ground state: since the embedded Hamiltonian only has real coefficients, the ground state is defined by ${4 \choose 2}=6$ real coefficients and the 1-RDM's non-diagonal vanishing entries impose 6 independent equations they must satisfy. Note that this does not happen for $N_c=2$ (70 coefficients but only 27 independent equations). Such 1-RDM states can be prepared by the minimal "Multireference" (MR) circuit shown in Fig.~\ref{fig:MR_circuit},  with $\theta=2\,\mathrm{arcsin}(\sqrt{n_0})$.

\subsection{Optimization}
On top of its low gate count, the MR circuit has the very useful property that it only comprises one parameter, carried by a rotation gate. As a consequence, its optimization only requires three energy measurements in virtue of the parameter-shift rule \cite{nakanishi_sequential_2020}, instead of a proper VQE optimization.

\subsection{NOization}
One apparent drawback of such a minimal circuit is that since it only produces diagonal 1-RDMs, it cannot be used in a NOization procedure. As an alternative, one could resort to a dressed Hamiltonian strategy \cite{mizukami_orbital_2020}, in which the transformation to the Natural Orbitals is determined by minimizing the expectation value of a Hamiltonian dressed with a variational orbital transformation whose parameters must be optimized on top of the parameter of the circuit. This strategy would also allow for the use of the small ansatz circuit that works in NO without adding further noise effects as the added computational burden only lies on the CPU, but would require full-fledged VQE with 17 parameters (one circuit parameters and sixteen matrix components for the Hamiltonian dressing). 

This would however not be advantageous over the strategy corresponding to previous work that consists in using the small HEA ansatz proposed in Ref.~\onlinecite{keen_quantum-classical_2020} as this latter ansatz has 3 CNOT gates too, and only 8 parameters.
Fortunately, one can show that the transformation to the NO basis
is always the same as the impurity model's parameters are varied,
so that it can be determined once and for all. We determine it by optimizing the HEA ansatz for $U=0, D=-0.4, \lambda_c=0.004$ (these embedding parameters being typical of those found along the RISB procedure for low $U$) with 10 sequential optimization cycles of the \textit{Rotosolve} algorithm \cite{ostaszewski_structure_2021} (that leverages the parameter-shift rule) and diagonalizing the 1-RDM associated with the optimal state (we test for 5 random initializations to limit the risk of dealing with an optimized state corresponding to a local minimum). We also test for the effect of a very simple error mitigation strategy, the zero-noise extrapolation \cite{he_zero-noise_2020}, here in the form of linear extrapolation with only 1 CNOT pair insertion. 

\subsection{Results}
We display the results obtained with the technique described above in the noise-free case as well as the noisy and noisy mitigated cases on Figure \ref{fig:Z_Nc1}. Results obtained with the hardware-efficient ansatz (HEA\cite{keen_quantum-classical_2020}) in presence of error mitigation are also plotted. 
The number of points in the Brillouin zone was set to 40 $\times$ 40 whereas the inverse temperature is $\beta=200$. The number of Nelder-Mead iterations is limited to 20.

Without noise, this strategy yields very accurate results, except for numerical instabilities at the Mott transition. When switching the noise on, the quasi-particle weight is at first underestimated but as $U$ grows, the error vanishes and the agreement with analytical data remains excellent for values of $U$ that are not too small. The poorer performances in the noisy regime at low $U$ are also observed at $N_c=2$. Here they correspond to a shift in the minimum of the optimization landscape that increases as the noise level grows, as can be seen on Figure \ref{fig:optim_ls_Nc1} (a)-(d). It may be due to the fact that the ansatz is overly complicated with regards to the state that must be prepared, so that noise has a greater impact. For instance here for $U=0$ the 3 CNOT gates are superfluous as the ground state corresponds to $\theta=0$, so that they add unnecessary noise. This intuition is supported by the fact that the noisy optimization of a 'product' ansatz, with a $RY$ gate applied on each qubit, better reproduces the noise-free, exact optimization landscape (see Fig.~\ref{fig:optim_ls_Nc1}(e)), and that an adverse effect of the parameter tuning method is ruled out (compare (d) and (e) in Fig.~\ref{fig:optim_ls_Nc1}). Overall, we observe a substantial enhancement in the accuracy of the results compared with the HEA ansatz, with a far more accurate rendering of the Mott transition. This comes with the additional benefit of necessitating a lot fewer shots, since the VQE optimization of the HEA circuit requires here to evaluate the cost function 240 times and is run 5 times to avoid local minima. 
The strength of this statement is only midly mitigated by the fact that the NO Hamiltonian has a greater number of Pauli terms, since the increase is only of a factor $7-8$ (from 7 terms to 52 for $U>0$). All in all, the gain in terms of shots can be evaluated as:

\begin{equation}
\frac{n_{\mathrm{evals}}^{\mathrm{HEA}}}{n_{{\mathrm{evals}}}^{\mathrm{MR}}} = \frac{5 \times 8 \times 10 \times 3 \times 7}{1 \times 1 \times 1 \times 3  \times 52} \simeq 54,
\end{equation}
which is a substantial advantage in the context of an effective computation on quantum hardware.

\bibliographystyle{apsrev4-1}
\bibliography {bib, bib2}

\end{document}